\documentclass[twocolumn,showpacs,preprintnumbers,nofootinbib,prd,superscriptaddress]{revtex4-1}
\usepackage{graphicx,amssymb,amsmath,amsthm,amsfonts,epsfig}

\usepackage[linktocpage]{hyperref}
\usepackage[usenames]{color}
\usepackage{epstopdf}

\def\nn{\nonumber}

\def\ii{{\rm i}}

\newcommand{\ben}{\begin{enumerate}}
\newcommand{\een}{\end{enumerate}}

\def\be{\begin{equation}}
\def\ee{\end{equation}}
\def\bea{\begin{eqnarray}}
\def\eea{\end{eqnarray}}
\newcommand{\beq}{\begin{eqnarray}}
\newcommand{\eeq}{\end{eqnarray}} 

\begin{document}
\title{Mixing of spherical and spheroidal modes in perturbed Kerr black holes}

\author{Emanuele Berti}\email{eberti@olemiss.edu}
\affiliation{Department of Physics and Astronomy, The University of Mississippi, University, MS 38677, USA.}

\author{Antoine Klein}\email{aklein@olemiss.edu}
\affiliation{Department of Physics and Astronomy, The University of Mississippi, University, MS 38677, USA.}

\date{\today} 

\begin{abstract} 
The angular dependence of the gravitational radiation emitted in
compact binary mergers and gravitational collapse is usually separated
using spin-weighted spherical harmonics ${}_sY_{\ell m}$ of spin
weight $s$, that reduce to the ordinary spherical harmonics $Y_{\ell
  m}$ when $s=0$. Teukolsky first showed that the perturbations of the
Kerr black hole that may be produced as a result of these events are
separable in terms of a different set of angular functions: the
spin-weighted {\em spheroidal} harmonics ${}_sS_{\ell m n}$, where $n$
denotes the ``overtone index'' of the corresponding Kerr quasinormal
mode frequency $\omega_{\ell m n}$. In this paper we compute the
complex-valued scalar products of the ${}_sS_{\ell m n}$'s with the
${}_sY_{\ell m}$'s (``spherical-spheroidal mixing coefficients'') and
with themselves (``spheroidal-spheroidal mixing coefficients'') as
functions of the dimensionless Kerr parameter $j$. Tables of these
coefficients and analytical fits of their dependence on $j$ are
available online for use in gravitational-wave source modeling and in
other applications of black-hole perturbation theory.
\end{abstract}

\maketitle

\section{Introduction}

Various angular functions, including scalar, vector, and tensor
spherical harmonics, are used to perform separation of variables in
the general relativity literature. These functions include the
Regge-Wheeler harmonics, the symmetric, trace-free tensors of Sachs
and Pirani, the Newman-Penrose spin-weighted spherical harmonics, and
the Mathews-Zerilli Clebsch-Gordan-coupled harmonics.  An excellent
review article by Thorne \cite{Thorne:1980ru} lists all of these
functions and discusses their mutual relations.

The spin-weighted spherical harmonics ${}_sY_{\ell m}$
\cite{Newman:1966ub,Goldberg:1966uu} are most commonly used to
separate the angular dependence of the gravitational radiation emitted
as a result of compact binary mergers and gravitational collapse in
numerical relativity simulations. Unfortunately, the ${}_sY_{\ell
  m}$'s are not ideal to study the perturbations of the rotating Kerr
black holes of mass $M$ and dimensionless angular momentum $j\equiv
a/M$ that may be formed as a result of compact binary mergers or
gravitational collapse (here and below $a$ is the usual Kerr
parameter, and we use geometrical units: $G=c=1$).

Teukolsky \cite{Teukolsky:1973ha,Press:1973zz} first realized that the
radiation produced by perturbed Kerr black holes is most conveniently
studied using a different set of angular functions: the spin-weighted
spheroidal harmonics ${}_sS_{\ell m}(a\omega)$ (henceforth SWSHs). The
differential equation defining these functions is a generalized
spheroidal wave equation \cite{Leaver:1986JMP}, and it results from
separating variables in the partial differential equations describing
the propagation of a spin-$s$ field in a rotating (Kerr) black hole
background. If we use the Kinnersley tetrad and Boyer-Lindquist
coordinates $(t,\,r,\,\theta,\,\phi)$, we assume a time dependence of
the form $e^{-\ii \omega t}$ and a $\phi$-dependence of the form
$e^{\ii m\phi}$, the SWSHs satisfy the equation
\cite{Teukolsky:1973ha,Leaver:1985ax}
\bea
\label{angularwaveeq}
&&\left [(1-x^2){}_sS_{\ell m,x} \right]_{,x} \\
&+&\left[
(cx)^2-2c sx+s+{}_{s}A_{\ell m}
 -{(m+sx)^2\over 1-x^2}
\right] {}_sS_{\ell m}=0\,,\nn
\eea
where $x\equiv \cos\theta$, $c\equiv a\omega$ and $\theta$ is the
Boyer-Lindquist polar angle.
The angular separation constant $_{s}A_{\ell m}$ is, in general,
complex. The spin-weight parameter takes on the values $s=0,\pm
1/2,\pm 1,\pm 2$ for massless scalar, spinor, vector and
tensor perturbations, respectively. 

\begin{figure*}[t]
\begin{center}
\begin{tabular}{ll}
\epsfig{file=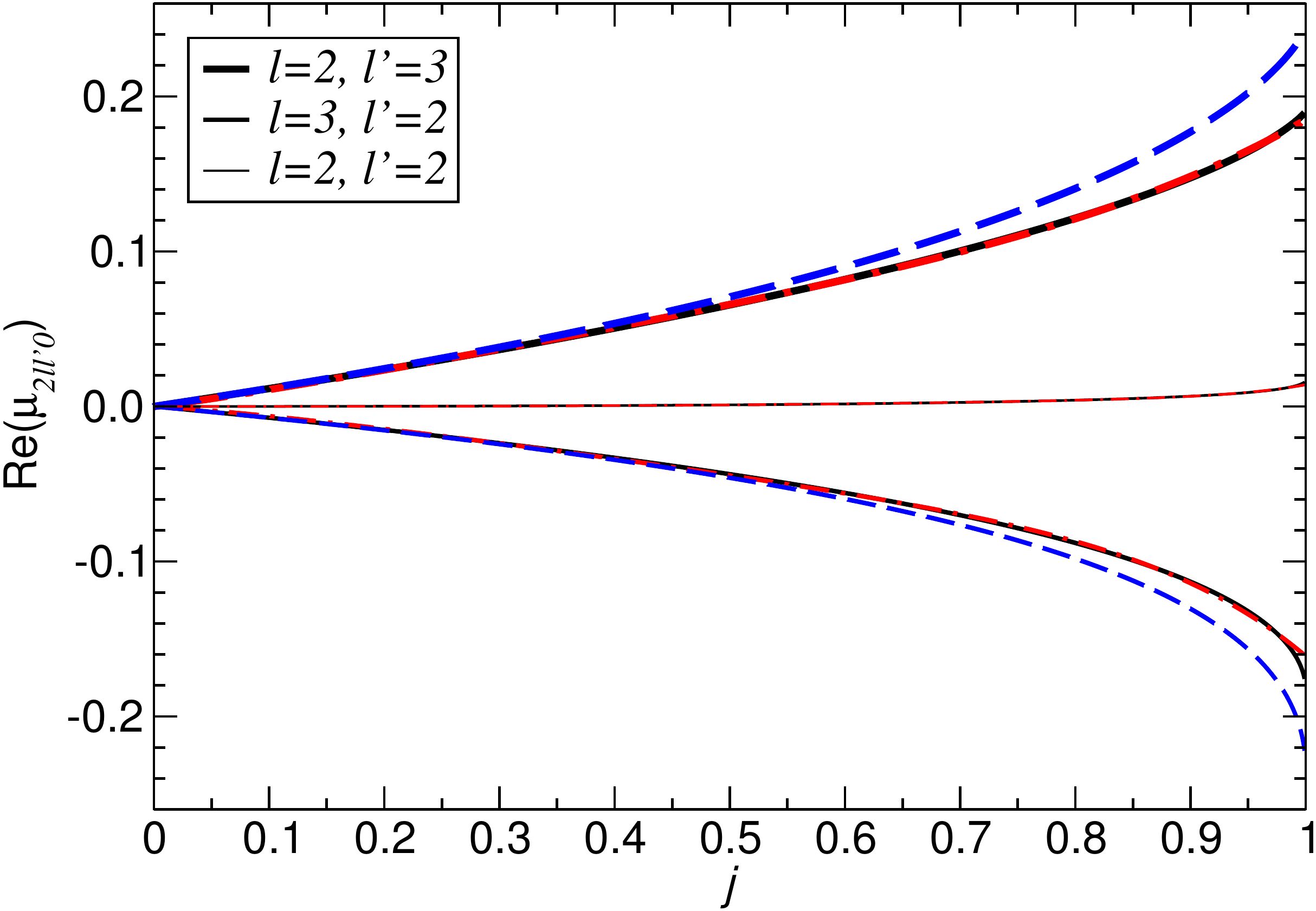,width=0.5\textwidth,angle=0,clip=true}&
\epsfig{file=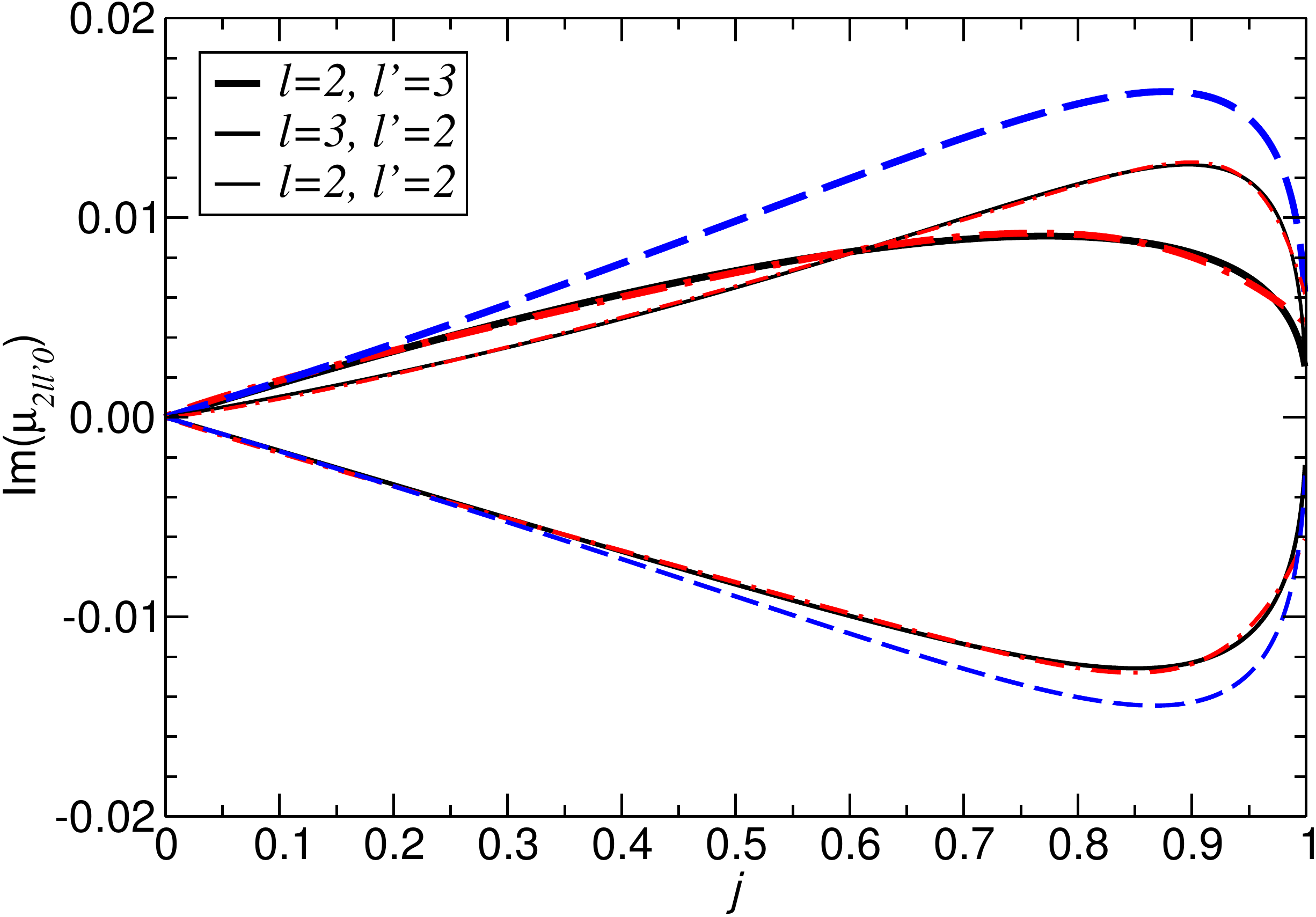,width=0.5\textwidth,angle=0,clip=true}\\
\end{tabular}
\caption{Real (left panel) and imaginary part (right panel) of the
  spherical-spheroidal mixing coefficients $\mu_{m\ell \ell' n'}$ for
  $m=2$, $n'=0$. Here we consider the dominant multipoles
  ($\ell=2,\,\ell'=3$) (very thick lines in the upper half of each
  panel), ($\ell=3,\,\ell'=2$) (thick lines in the lower half of each
  panel) and ($\ell=\ell'=2$) (thin lines in each panel). Solid lines
  (black online) correspond to the numerical calculation presented in
  this paper; dash-dotted lines (red online, almost indistinguishable
  from the black lines) are the fitting relations of
  Eq.~(\ref{mufit}); dashed lines (blue online) are the leading-order
  Press-Teukolsky approximation \cite{Press:1973zz}.  In the case
  $\ell=\ell'=2$ the mixing coefficients are close to unity, so we
  actually plot $1-\mu_{2 2 2 0}$; furthermore, to leading order the
  Press-Teukolsky calculation predicts that they are exactly equal to
  unity, so their approximation is not shown in the plot.
\label{fig:m2n0lowest}}
\end{center}
\end{figure*}

When $s=0$ the SWSHs reduce to the ordinary (scalar) spheroidal wave
functions \cite{flammer}.
In the limit $c\to 0$ (corresponding to the Schwarzschild limit) the
spin-weighted spheroidal harmonics reduce to spin-weighted spherical
harmonics ${}_sY_{\ell m}$ \cite{Newman:1966ub,Goldberg:1966uu}, for
which
\begin{equation}\label{schwlim}
{}_sA_{\ell m}=\ell(\ell+1)-s(s+1)\,.
\end{equation}
and
\be
\int~{_{-2}Y_{\ell m}}^*~ _{-2}Y_{\ell 'm'} d\Omega=\delta_{\ell ,\ell'} \delta_{m,m'}\,.
\ee
The ordinary spherical harmonics are spin-weighted spherical harmonics
with $s=0$.

The gravitational waves emitted by newly formed Kerr black holes can
be decomposed as a superposition of complex quasinormal modes (QNMs)
with frequencies $\omega_{\ell mn}$, where the ``overtone index'' $n$
measures the magnitude of the imaginary part of the frequencies:
low-$n$ modes damp most slowly, and therefore they dominate the
response of the black hole
\cite{Kokkotas:1999bd,Berti:2009kk,Konoplya:2011qq}.
Each QNM can be associated to a SWSH angular eigenfunction
${}_sS_{\ell m n}\equiv {}_sS_{\ell m}(a\omega_{\ell mn})$ labeled by
the corresponding overtone index $n$
\cite{Leaver:1985ax,Leaver:1986JMP,Berti:2005ys}.
Due to their importance in black-hole physics, the properties of SWSHs
have been investigated in some depth
\cite{Teukolsky:1973ha,Press:1973zz,Breuer:1975,Fackerell:1977,Seidel:1988ue,Berti:2005gp}.
Press and Teukolsky \cite{Press:1973zz} provided a polynomial fit in
$c$ of the eigenvalues ${}_sA_{\ell m}$, which is valid up to $c \sim 3$.
A formal perturbation expansion in powers of $c$ was carried out by
Fackerell and Crossman \cite{Fackerell:1977} (see also
\cite{Seidel:1988ue}, where some typos were corrected). Analytic
expansions for small and large values of $c$ were discussed and
compared to numerical results in \cite{Berti:2005gp}.

In practice, only the first few QNMs contribute noticeably to the
ringdown radiation from a newly formed Kerr black hole. These modes
were first investigated in detail by Leaver and Onozawa
\cite{Leaver:1985ax,Onozawa:1996ux}. Higher-order modes may have some
relevance in the context of black-hole thermodynamics and quantum
gravity (see
\cite{Berti:2003zu,Berti:2003jh,Neitzke:2003mz,Berti:2004um,Hod:2005ha,Keshet:2007nv,Keshet:2007be,Kao:2008sv}),
but they will not be discussed in this paper.

The main motivation for the present study is that the use of spherical
harmonics (rather than SWSHs) induces significant {\em mode mixing} in
numerical relativity simulations of black-hole binary mergers. This
mixing is particularly evident in the $(\ell=3,\,m=2)$ spin-weighted
spherical harmonic mode, where (as first noticed in
\cite{Buonanno:2006ui}) the ringdown radiation is a superposition of
the $\omega_{320}$ and $\omega_{220}$ modes. Subsequent studies
confirmed this finding
\cite{Berti:2007fi,Schnittman:2007ij,Baker:2008mj,Kelly:2011bp,Pan:2011gk},
and it was recently proved beyond any reasonable doubt that the
observed QNM mixing occurs because spherical harmonics contain a
superposition of several spheroidal harmonics
\cite{Kelly:2012nd,London:2014cma,Taracchini:2014zpa}.

Mathematically, mode mixing occurs because, to leading order in
perturbation theory, SWSHs with angular indices $(\ell,\,m)$ are a
superposition of spherical harmonics with the {\em same} value of $m$
but different values of $\ell'\neq \ell$. As shown by Press and
Teukolsky \cite{Press:1973zz},
\be\label{pert-th}
{}_sS_{\ell m}=
{}_sY_{\ell m}+\sum_{\ell'\neq \ell}
\frac{\langle s \ell' m|\mathfrak{h}_1|s \ell m\rangle}{\ell(\ell+1)-\ell'(\ell'+1)}
{}_sY_{\ell' m}+\dots
\ee
where the specific form of $\langle s \ell' m|\mathfrak{h}_1|s \ell
m\rangle$ is not important for the moment (cf. Section \ref{sec:PT}
below for details).

A systematic investigation of the mixing between spherical and
spheroidal harmonics is needed to construct semianalytical models
of the transition from merger to ringdown, both in the extreme
mass-ratio limit \cite{Taracchini:2014zpa,Harms:2014dqa} and for
comparable-mass binaries
\cite{Damour:2014yha,Healy:2014eua}. Furthermore, a better
understanding of this mixing can help in selecting the optimal frame
to analyze generic precessing black-hole binary mergers
\cite{Gualtieri:2008ux,Campanelli:2008nk,O'Shaughnessy:2011fx,Boyle:2011gg,Boyle:2013nka}. More
in general, a ``dictionary'' relating spherical and spheroidal modes
is useful in all applications of black-hole perturbation theory.

Quite surprisingly (and to the best of our knowledge) no systematic
investigation of mode mixing is available in the literature. The main
goal of this paper is to fill this gap by computing the complex
universal functions $\mu_{m\ell \ell'n'}(j)$ of the dimensionless
black-hole spin $j\equiv a/M\in [0,\,1]$ defined by the following
inner product:
\be \int {_{s}S_{\ell 'm'n'}^* ~ _{s}Y_{\ell m}}\,d\Omega=\mu_{m\ell \ell'n'}(j)
\delta_{m,m'}\,, \label{mcoeff}
\ee
where $s=-2$, $-1$ or $0$, and the Kronecker symbol $\delta_{m,m'}$
comes from the $e^{\ii m \phi}$-dependence of the harmonics.

Another goal of this paper is to produce a catalog of the following
quantities, that are of interest for ringdown data analysis in the
context of gravitational-wave detection
\cite{Berti:2005ys,Berti:2007zu}:
\be \int {_{-2}S_{\ell' m' n'}}^*~
_{-2}S_{\ell m n}\, d\Omega=\alpha_{m\ell \ell'nn'}(j)
\delta_{m,m'}\,. \label{acoeff}
\ee
The functions $\alpha_{m\ell \ell'nn'}(j)$ were evaluated numerically
for specific values of the indices and for a single value of the spin
parameter ($j=0.98$ in Table I, and $j=0.8$ in Tables II and III) in
\cite{Berti:2005gp}. Here we extend that calculation to all dominant
modes and to all values of $j\in [0,\,1]$. Our numerical results for
both sets of coefficients $\mu_{m\ell \ell'n'}(j)$ (henceforth the
{\em spherical-spheroidal mixing coefficients}) and $\alpha_{m\ell
  \ell'nn'}(j)$ (henceforth the {\em spheroidal-spheroidal mixing
  coefficients}) are available online \cite{rdweb}.

In Fig.~\ref{fig:m2n0lowest} we illustrate the importance of going
beyond the Press-Teukolsky perturbation-theory calculation in
computing the mixing coefficients. There we consider the fundamental
($n'=0$) QNM with $m=2$ and we plot $\mu_{m\ell \ell'n'}(j)$ for
$\ell=2,\,3$, $\ell'=2,\,3$, i.e. for the dominant multipoles in
binary black-hole mergers.  The plot compares: (1) the numerical
calculation of the coefficients $\mu_{2\ell \ell'0}$ reported in this
paper, (2) a power-law fit to the numerical results
[cf. Eq.~(\ref{mufit}) below], and (3) the approximate value of these
coefficients predicted by the Press-Teukolsky expansion of
Eq.~(\ref{pert-th}). Fig.~\ref{fig:m2n0lowest} shows that the
Press-Teukolsky approximation is adequate for small spins, but it is
not accurate enough for fast rotating black holes, with relative
errors\footnote{Analyzing the analytical predictions for different
  $\mu_{m \ell \ell' n'}$ at the maximum spin value considered here
  ($j=0.999$), we find that (i) the absolute deviations from the
  analytical prediction in the mixing coefficients for counterrotating
  modes ($m < 0$) when $\ell = \ell'$ can be of order unity for large
  $n'$, and they are consistently in excess of $0.5$ for $n' \geq 3$;
  (ii) the relative deviations for $m>0$ and $\ell \neq \ell'$ are
  usually larger than $10\%$: e.g. they are between $27\%$ and $29\%$
  for $\mu_{223n'}$, irrespective of $n'$.} of order $\sim 30\%$ when
$j\to 1$ even for $n' = 0$.

The outline of the paper is as follows. We first recall some
properties of the SWSHs (Section \ref{sec:SWSHs}). Then we show the
results of our numerical calculation of the mixing coefficients and we
give analytical fits of the $j$-dependence of the cofficients (Section
\ref{sec:mix}). In the conclusions we point out possible applications
of this calculation and directions for future work.

\section{Spin-weighted spheroidal harmonics}\label{sec:SWSHs}

Leaver \cite{Leaver:1985ax} found the following series solution of the
SWSH equation (\ref{angularwaveeq}):
\begin{equation}\label{leavers}
{}_sS_{\ell m n}(\theta,\phi)=e^{i m \phi}e^{c_{\ell mn} x}\left(1+x\right)^{k_-}\left(1-x\right)^{k_+}
\sum_{p=0}^\infty a_p(1+x)^p\,,
\end{equation}
where $k_{\pm}\equiv |m\pm s|/2$, and $x=\cos\theta$. The expansion
coefficients $a_p$ are obtained from a three-term recursion relation
that can be found, e.g., in \cite{Leaver:1985ax,Berti:2005gp}. 

The angular separation constant ${}_{s}A_{\ell m n}$, $c_{\ell
  mn}=a\omega_{\ell mn}$ and the SWSHs ${}_sS_{\ell m n}$ are, in
general, complex. They take on real values only in the oblate case
($c_{\ell m n}\in \mathbb{R}$) or, alternatively, in the prolate case
($c_{\ell m n}$ pure-imaginary) with $s=0$. Some useful symmetry
properties hold (see eg. \cite{Leaver:1985ax}):

\begin{itemize}
\item[(i)] Given eigenvalues for (say) $m>0$, those for $m<0$
are readily obtained by complex conjugation:
\begin{equation}\label{conj}
{}_sA_{\ell mn}={}_sA_{\ell -mn}^*\,;
\end{equation}

\item[(ii)] Given eigenvalues for (say) $s<0$, those for $s>0$
are given by
\begin{equation}\label{negs}
_{-s}A_{\ell mn}={}_sA_{\ell mn}+2s\,.
\end{equation}
Exploiting these symmetries, in our numerical calculations we only
consider $s\leq 0$ and $m\geq 0$. In practice this means that we only
compute the positive-frequency QNMs, even though each mode consists
of {\em both} a positive-frequency and a negative-frequency component:
see \cite{Leaver:1985ax,Berti:2005ys} for more extensive discussions.

\item[(iii)] Let us define $\rho_{\ell mn}\equiv i c_{\ell mn}$. If
  $\rho_{\ell mn}$ and $_{-s}A_{\ell mn}$ correspond to a solution for
  given $(s,\,l,\,m,\,n)$, then another solution can be obtained by
  the following replacements: $m\to -m$, $\rho_{\ell mn}\to \rho_{\ell
    mn}^*$, $_{-s}A_{\ell mn}\to _{-s}A^*_{\ell -mn}
$.
\end{itemize}

Leaver's solution gives a simple and practical algorithm for the
numerical calculation of eigenvalues ${}_sA_{\ell mn}$ and
eigenfunctions ${}_sS_{\ell m n}$ for a perturbed Kerr black hole. The
procedure we use is standard and it is described in many papers
\cite{Leaver:1985ax,Onozawa:1996ux,Berti:2004md,Berti:2009kk}, so here
we give a very concise summary. Start from the analytically known
angular eigenvalue for a given overtone $n$ in the Schwarzschild
limit, Eq.~(\ref{schwlim}). In the Kerr space-time, linear
gravitational perturbations are described by a pair of coupled
differential equations: one for the angular part of the perturbations,
and the other for the radial part.  The radial equation is given,
e.g., in \cite{Teukolsky:1973ha,Leaver:1985ax}. The angular equation
is the SWSH equation (\ref{angularwaveeq}). Boundary conditions for
the two equations can be cast as a pair of three-term continued
fraction relations. Solve the radial continued-fraction equation to
find $\omega_{\ell mn}$ in the Schwarzschild limit. Now increase $j$
in small increments and, for given values of $(s,\,\ell,\,m,\,n)$,
look for simultaneous zeros of the radial and angular continued
fraction equations to find both the ``radial eigenvalue''
$\omega_{\ell mn}$ and the angular separation constant $_{s}A_{\ell
  mn}$, using the values computed for smaller $j$ as initial guesses in
the numerical search.  Once the radial and angular eigenvalues are
known, the series coefficients $a_p$ can be computed using the
recursion relation and plugged into the series solution
(\ref{leavers}) to get the corresponding eigenfunction to the required
precision. In our numerical calculations we truncate the series at
some $p=p_{\rm max}$ such that the inclusion of subsequent terms would
not modify the series by more than one part in $10^6$.  This algorithm
only determines the eigenfunction up to a normalization constant,
which can easily be fixed by imposing the normalization condition
\be
\int |{}_sS_{\ell m n}|^2 d\Omega=1\,.
\ee

\begin{figure*}[t]
\begin{center}
\begin{tabular}{ll}
\epsfig{file=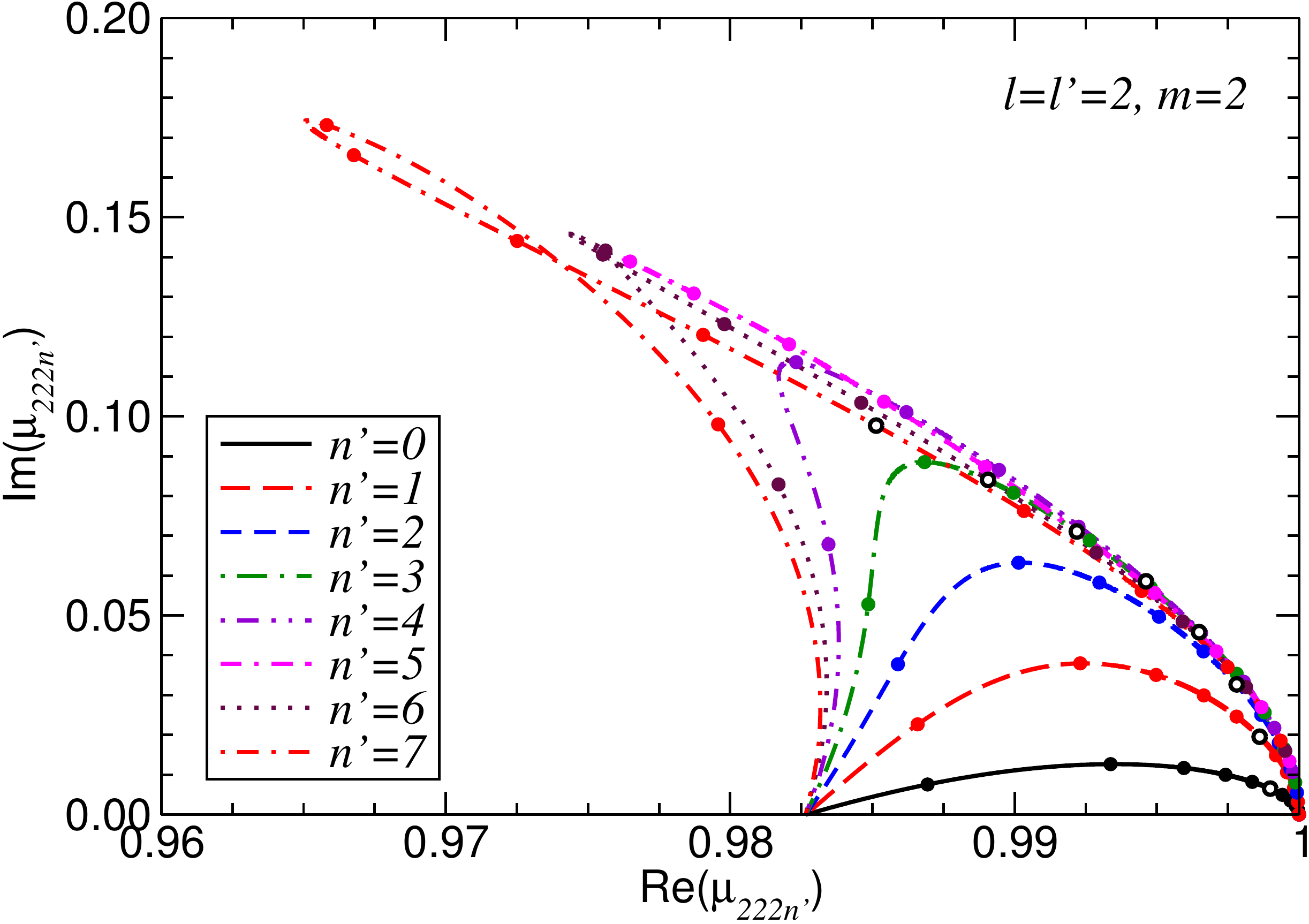,width=0.5\textwidth,angle=0,clip=true}&
\epsfig{file=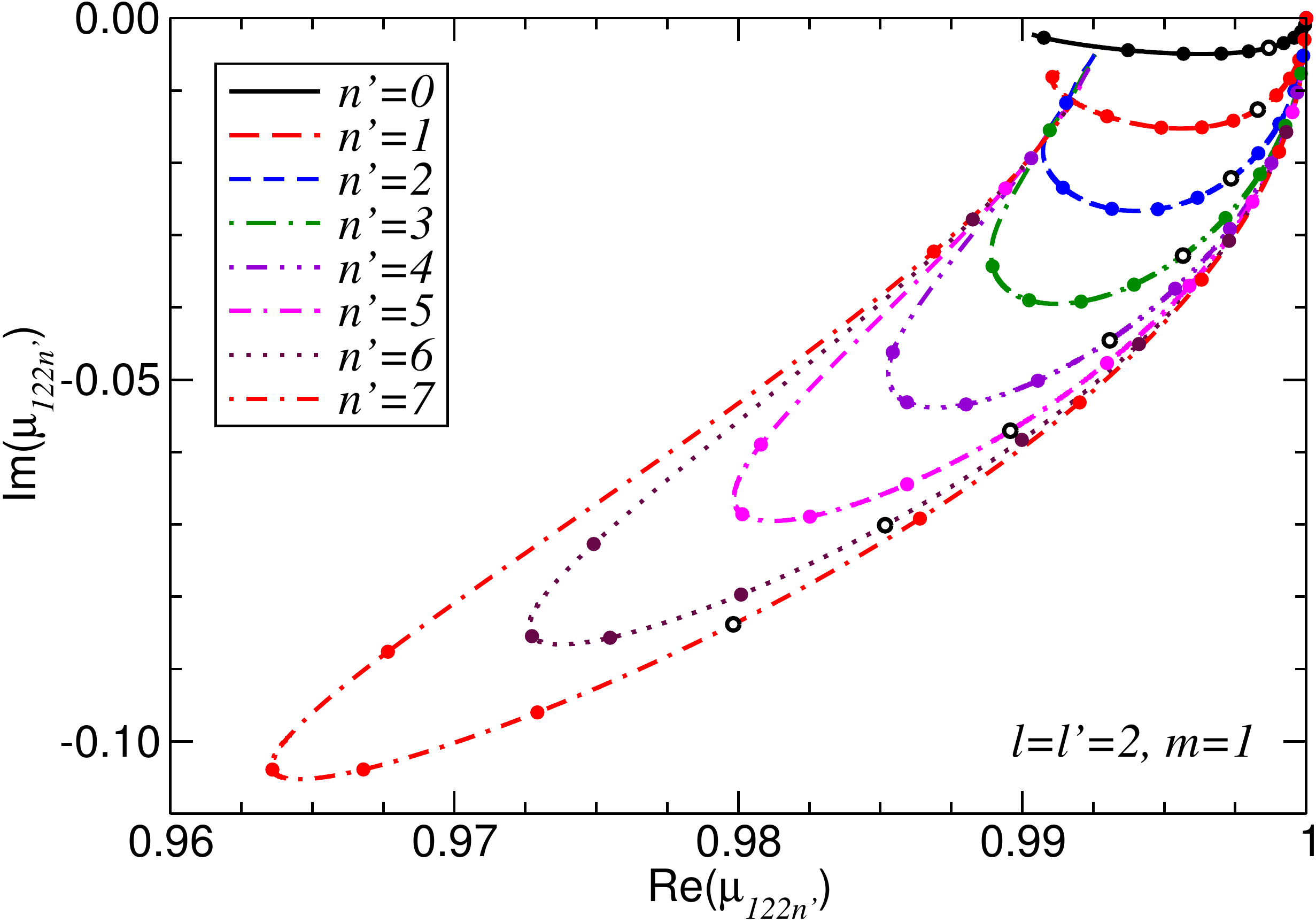,width=0.5\textwidth,angle=0,clip=true}\\
\end{tabular}
\caption{Trajectories traced by the mixing coefficients $\mu_{m\ell
    \ell' n'}$ with $\ell=\ell'=2$, $m=2$ (left panel) and $m=1$
  (right panel) as the Kerr parameter increases from $j=0$ (where
  $\mu_{m22 n'}=1$) to the nearly extremal Kerr limit. Each
  curve can be thought of as a parametric plot, where the parameter is
  $j$. Filled circles denote the following discrete values of $j$:
  $j=0,\,0.1,\,0.2,\,\dots,\,0.9,0.99$. To guide the eye, along each
  trajectory the dimensionless Kerr parameter $j=0.5$ is denoted by a
  hollow circle.
\label{fig:l2lp2xmg}}
\end{center}
\end{figure*}
\begin{figure*}[t]
\begin{center}
\begin{tabular}{ll}
\epsfig{file=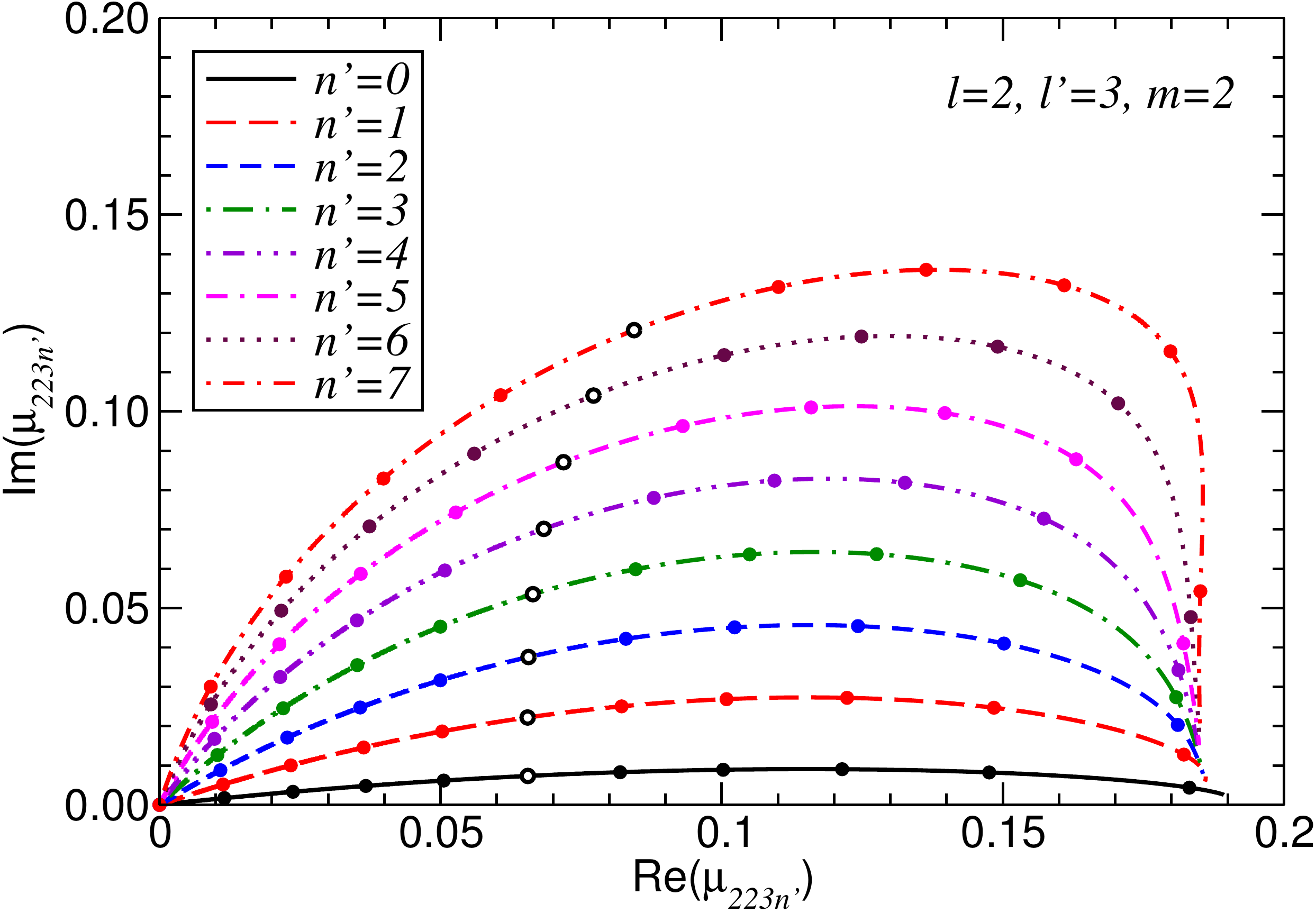,width=0.5\textwidth,angle=0,clip=true}&
\epsfig{file=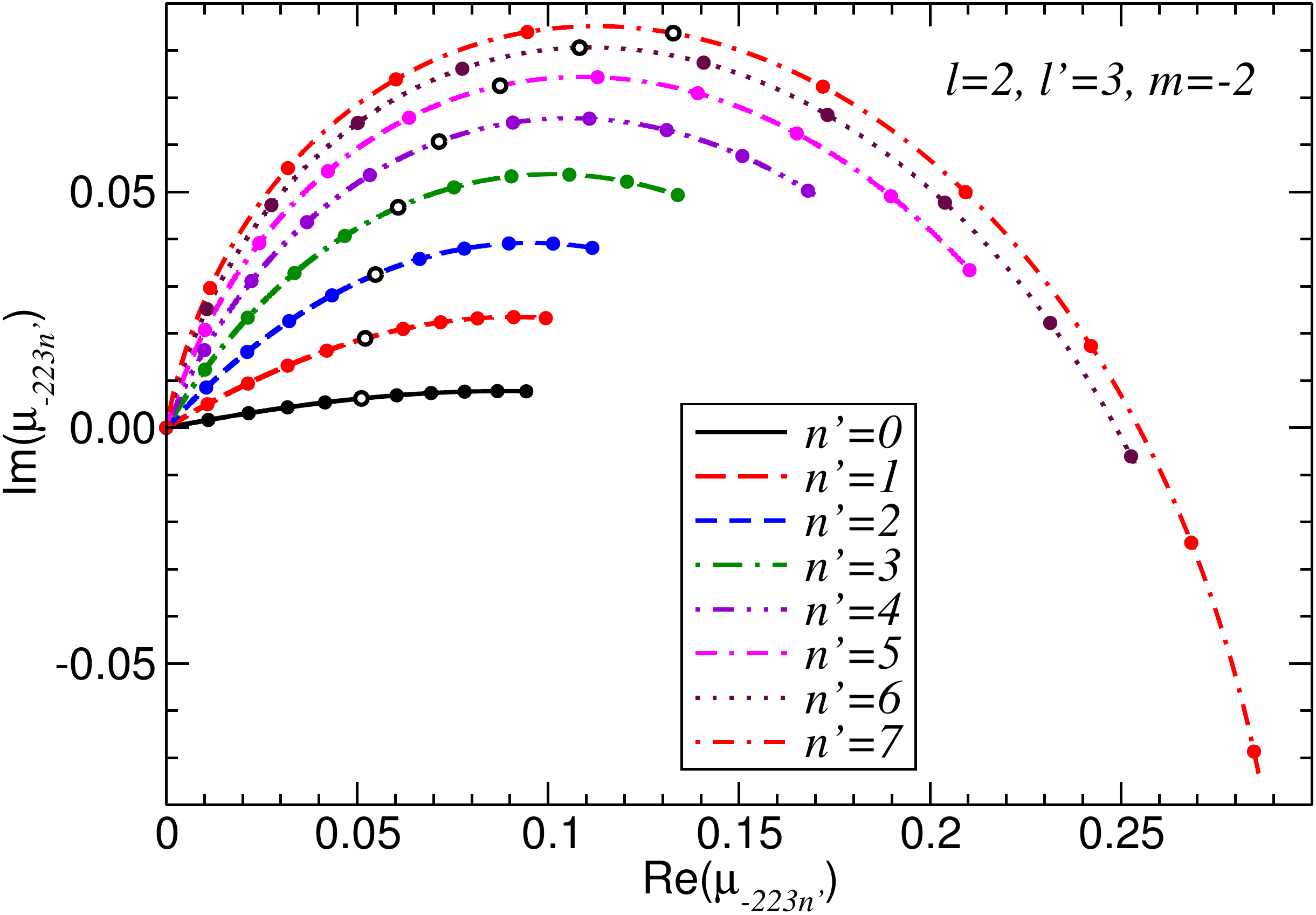,width=0.5\textwidth,angle=0,clip=true}\\
\epsfig{file=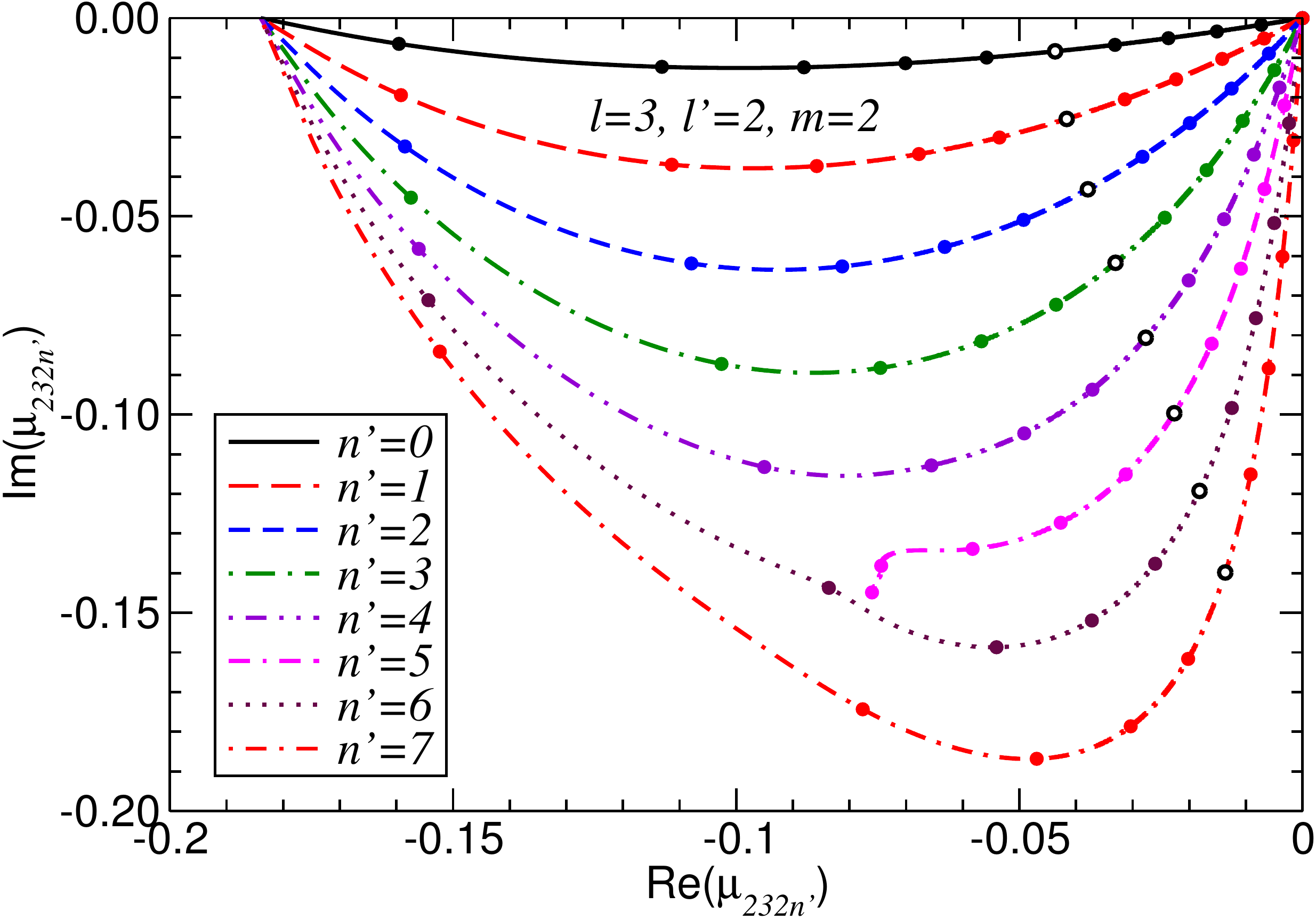,width=0.5\textwidth,angle=0,clip=true}&
\epsfig{file=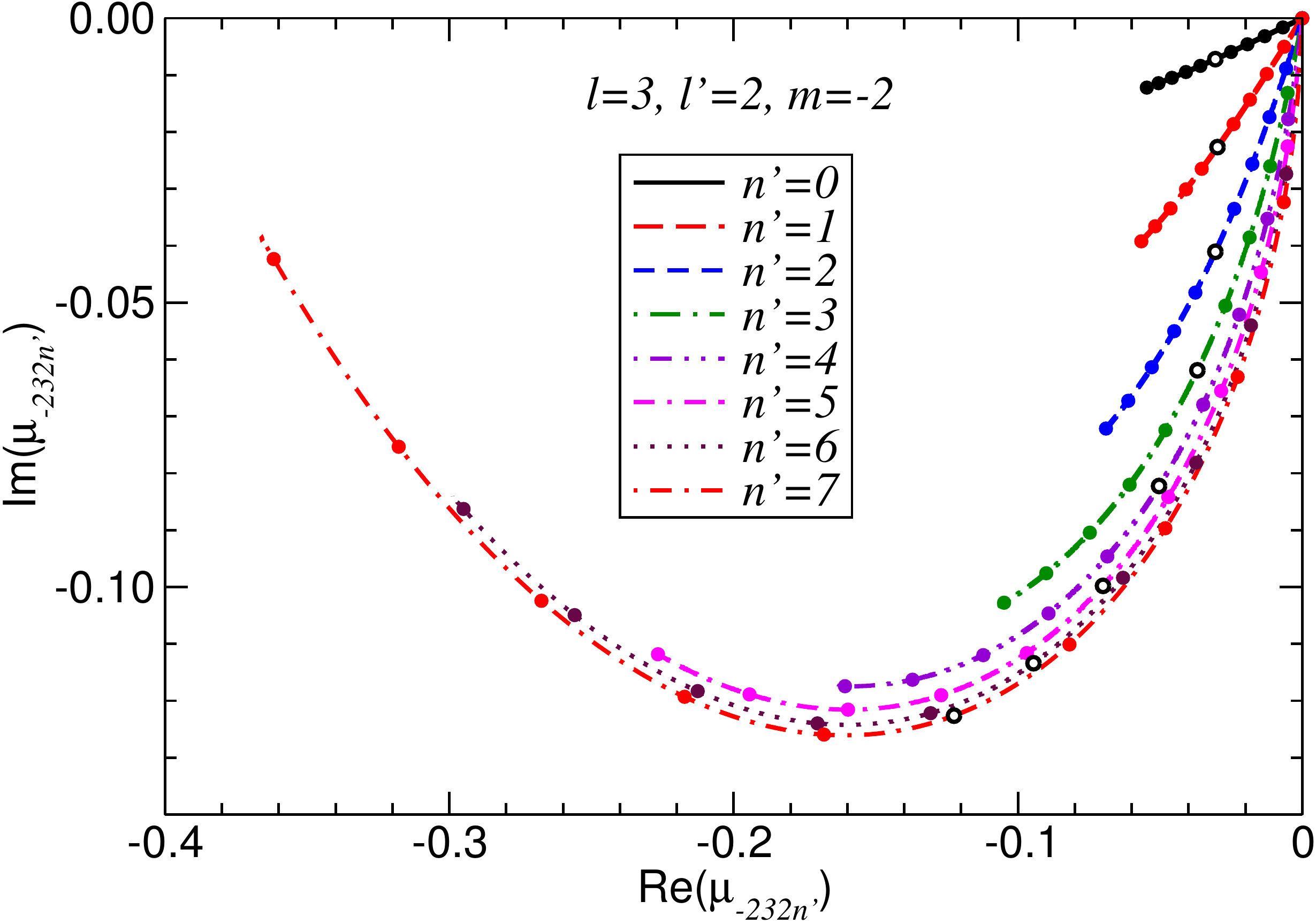,width=0.5\textwidth,angle=0,clip=true}\\
\end{tabular}
\caption{Trajectories traced by the mixing coefficients $\mu_{m\ell
    \ell' n'}$ in the complex plane as the Kerr parameter increases
  from $j=0$ (where $\mu_{m\ell \ell' n'}=0$ for $\ell\neq \ell'$) to
  $j=1$.  Panels in the top row refer to $(\ell,\,\ell')=(2,\,3)$,
  those in the bottom row to $(\ell,\,\ell')=(3,\,2)$; left panels are
  for $m=2$, right panels for $m=-2$.
\label{fig:l2l3m2mm2}}
\end{center}
\end{figure*}
\begin{figure}[t]
\begin{center}
\epsfig{file=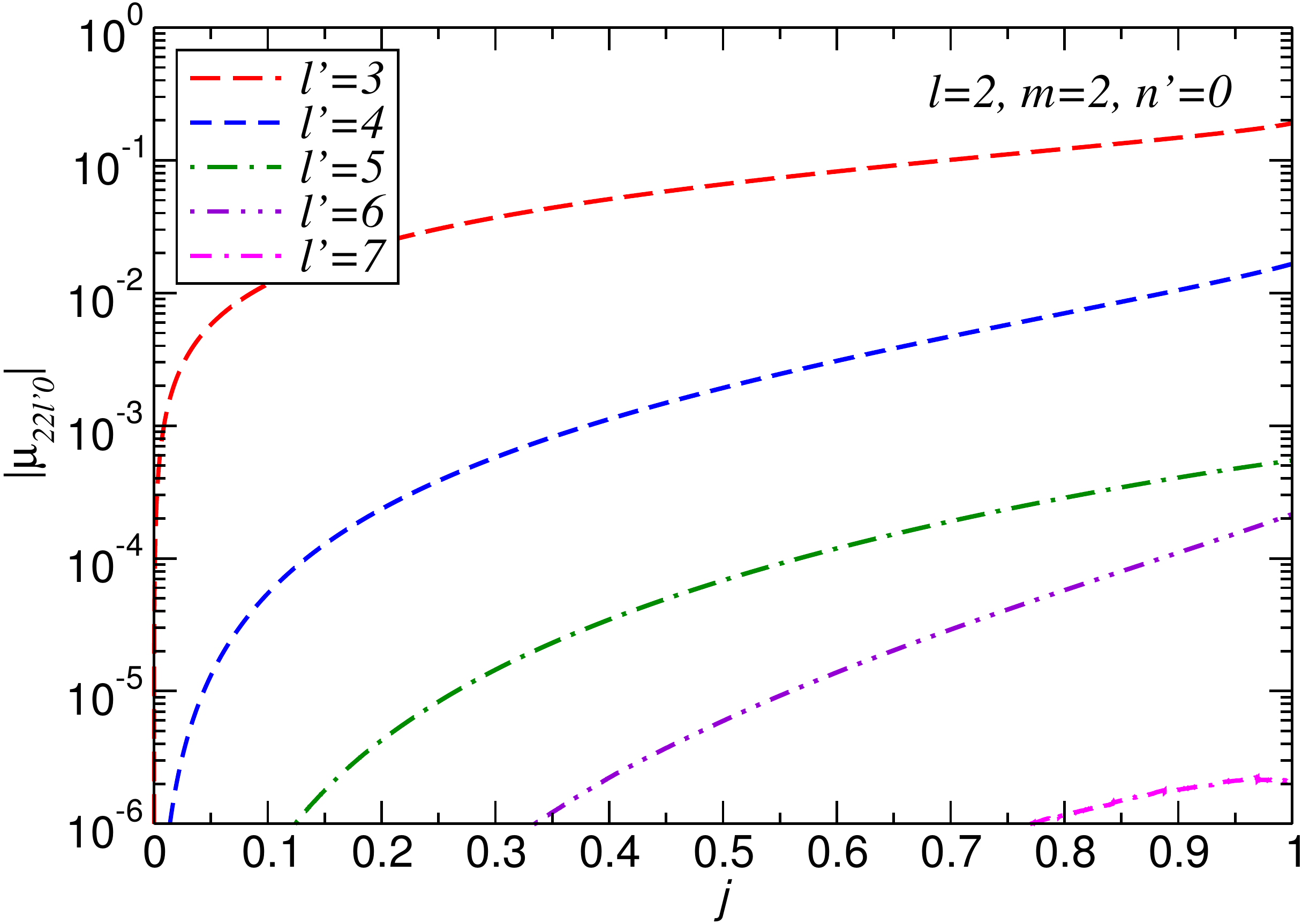,width=0.5\textwidth,angle=0,clip=true}
\caption{Absolute values of the mixing coefficients $|\mu_{m\ell \ell'
    n'}|$ with $\ell=m=2$, $n'=0$ and different values of
  $\ell'=3,\,\dots,7$, illustrating the roughly exponential decay of
  the mixing coefficients with $|\ell'-\ell|$.\label{fig:l2lpn}}
\end{center}
\end{figure}
%

\section{Mixing coefficients}\label{sec:mix}

In this section we present and discuss our numerical results for both,
the spherical-spheroidal mixing coefficients $\mu_{m\ell \ell'n'}(j)$
and the spheroidal-spheroidal mixing coefficients $\alpha_{m\ell
  \ell'nn'}(j)$. We also present power-law fits of the dependence of
these coefficients on the dimensionless Kerr parameter $j$.

\subsection{The spherical-spheroidal mixing coefficients}

Fig.~\ref{fig:l2lp2xmg} shows how the mixing coefficients for
$\ell=\ell'=2$ and $m=2$ (left) or $m=1$ (right) behave for the first
8 QNMs ($n'=0,\,\dots,\,7$) as the Kerr parameter increases from the
Schwarzschild limit $j=0$ (where $\mu_{m\ell \ell' n'}=1$) to the
extremal Kerr limit $j=1$. Each curve can be thought of as a
parametric plot, where the parameter along the curve is $j$. Circles
denote the following discrete values of $j$:
$j=0,\,0.1,\,0.2,\,\dots,\,0.9,0.99$. The numerical data are truncated
at $j=0.999$, because the behavior of QNMs for values of $j$ very
close to unity requires a special treatment
\cite{Yang:2012pj,Yang:2013uba}.

As first shown by Detweiler, for corotating modes with $\ell=m$ the
imaginary part of the quasinormal frequencies goes to zero as $j\to 1$
\cite{Detweiler:1980gk}. The physical reason for this behavior is that
QNMs can be thought of as perturbations of null geodesics
\cite{Mashhoon:1985cya,Cardoso:2008bp,Yang:2012he,Yang:2012pj,Yang:2013uba}. 
In the extremal limit the spherical photon orbit approaches the
horizon and the frequency of most QNMs with $\ell=m$ becomes equal to
$m\Omega_{\rm H}$, where $\Omega_{\rm H}=a/(2M r_+)$ is the angular
velocity and $r_+=M+\sqrt{M^2-a^2}$ is the Boyer-Lindquist radius of
the (outer) horizon. Whenever the QNM frequency tends to the critical
value for superradiance $m\Omega_{\rm H}$ the black hole becomes
marginally unstable, the eigenvalues of the SWSHs become real, and the
SWSHs themselves become oblate in the language of Flammer's monograph
\cite{flammer}. A surprising exception to this rule is the overtone
with $n'=5$: this oddity was first noticed by Onozawa (cf. Fig.~4 of
\cite{Onozawa:1996ux}). As a consequence, the mixing coefficient
corresponding to the mode with $n'=5$ in the left panel of
Fig.~\ref{fig:l2lp2xmg} is also exceptional, and it does not ``turn
around'' to meet the other modes on the real axis as $j\to 1$.

Fig.~\ref{fig:l2l3m2mm2} shows the dominant mixing coefficients for
the first 8 QNMs ($n'=0,\,\dots,\,7$) with $(\ell,\,\ell')=(2,\,3)$,
$(\ell,\,\ell')=(3,\,2)$ and $m=2$ or $m=-2$. We choose to display
these particular values of the mixing coefficients because they are
the most relevant to explain the spherical-spheroidal mode mixing
studied in \cite{Kelly:2012nd,London:2014cma} (for the $m=2$ modes of
comparable mass black-hole mergers) and \cite{Taracchini:2014zpa} (for
the $m=\pm 2$ modes of extreme-mass-ratio black-hole mergers). Once
again, note that the inner product becomes purely real near the
superradiant frequency for modes with $m=2$, because the imaginary
part of the QNM frequencies with $\ell=m$ tends to zero and the
harmonics become oblate -- the overtone with $n'=5$ being, again, the
exception. The plot also highlights the fact that the absolute value
of the mixing coefficients is typically larger for large spins (at
fixed overtone number $n'$) and for large overtone numbers (at fixed
spin $j$).

In Fig.~\ref{fig:l2lpn} we plot the absolute value of the mixing
coefficients $|\mu_{m\ell \ell' n'}|$ with $\ell=2$, $m=2$, $n'=0$ as
$\ell'$ increases. The figure shows that (perhaps unsurprisingly) mode
coupling decays roughly exponentially with $|\ell' - \ell|$.

Numerical tables of $\mu_{m\ell \ell' n'}(j)$ for all modes with
$|s|\leq \ell \leq 7$, $-\ell \leq m \leq \ell$, $-\ell' \leq m \leq
\ell'$, $0\leq n'\leq 7$ for $s=-2$, and $0\leq n'\leq 3$ for $s=-1$
and $s=0$ can be found online \cite{rdweb}.

\subsection{The spheroidal-spheroidal mixing coefficients}

\begin{figure}[t]
\begin{center}
\epsfig{file=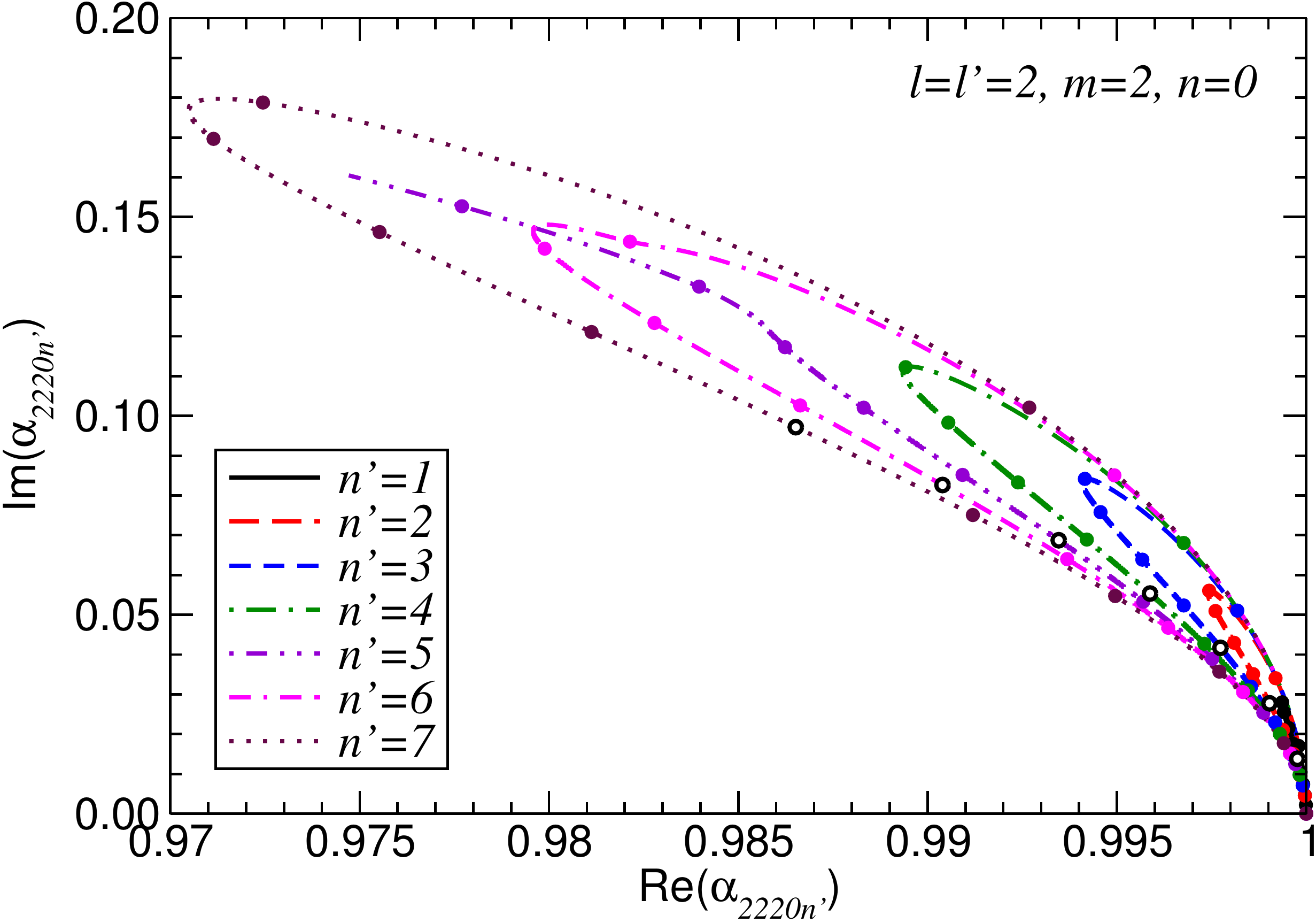,width=0.5\textwidth,angle=0,clip=true}
\caption{Trajectories described in the complex plane by the
  spheroidal-spheroidal mixing coefficients $\alpha_{m\ell
    \ell'nn'}(j)$ with $\ell=\ell'=m=2$, $n=0$ and different values of
  the {\em overtone index} $n'\geq 1$ as the dimensionless spin
  increases from $j=0$ to $j=1$.}
\label{fig:spheroidalsn07}
\end{center}
\end{figure}
\begin{figure*}[t]
\begin{center}
\begin{tabular}{ll}
\epsfig{file=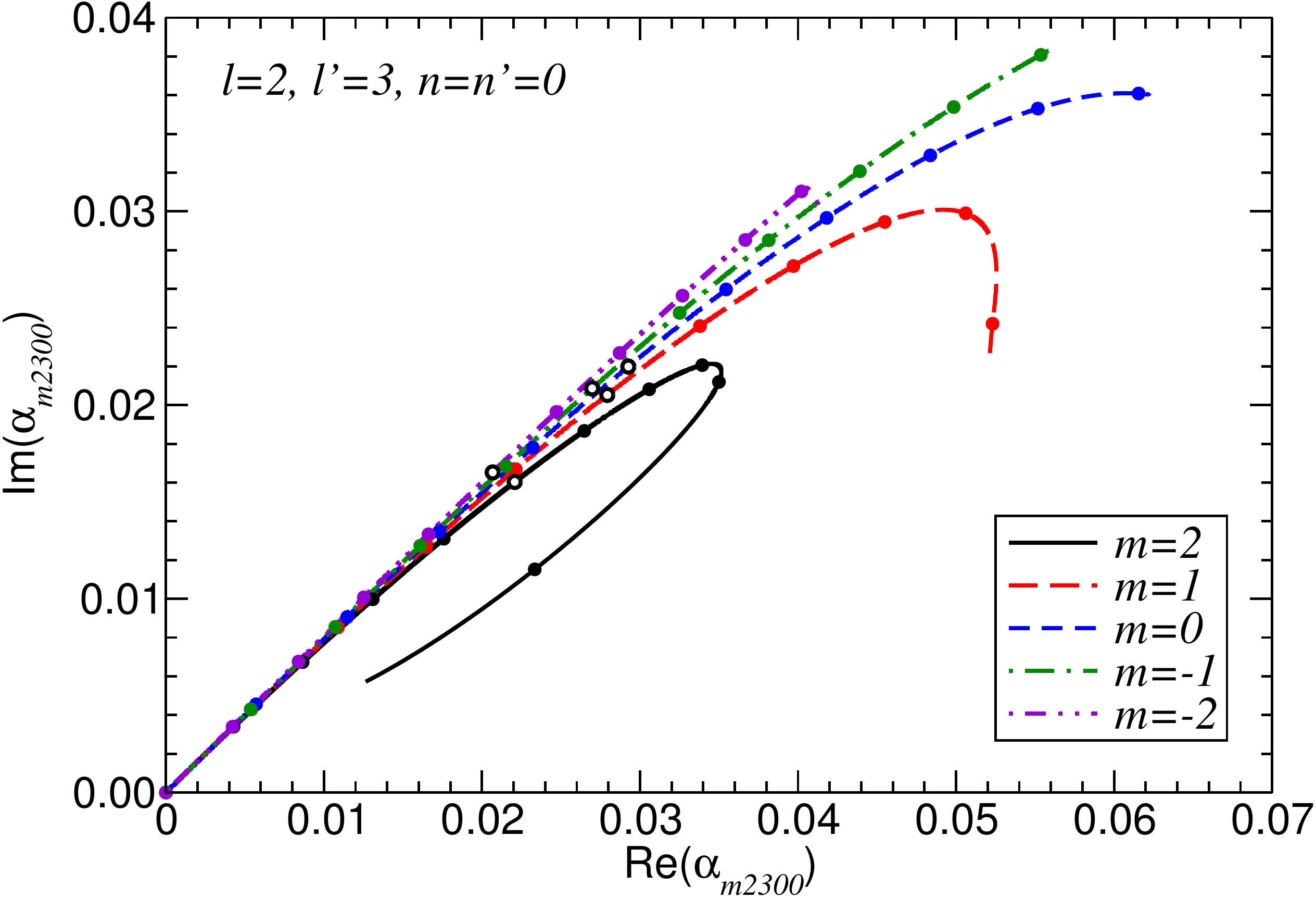,width=0.5\textwidth,angle=0,clip=true}&
\epsfig{file=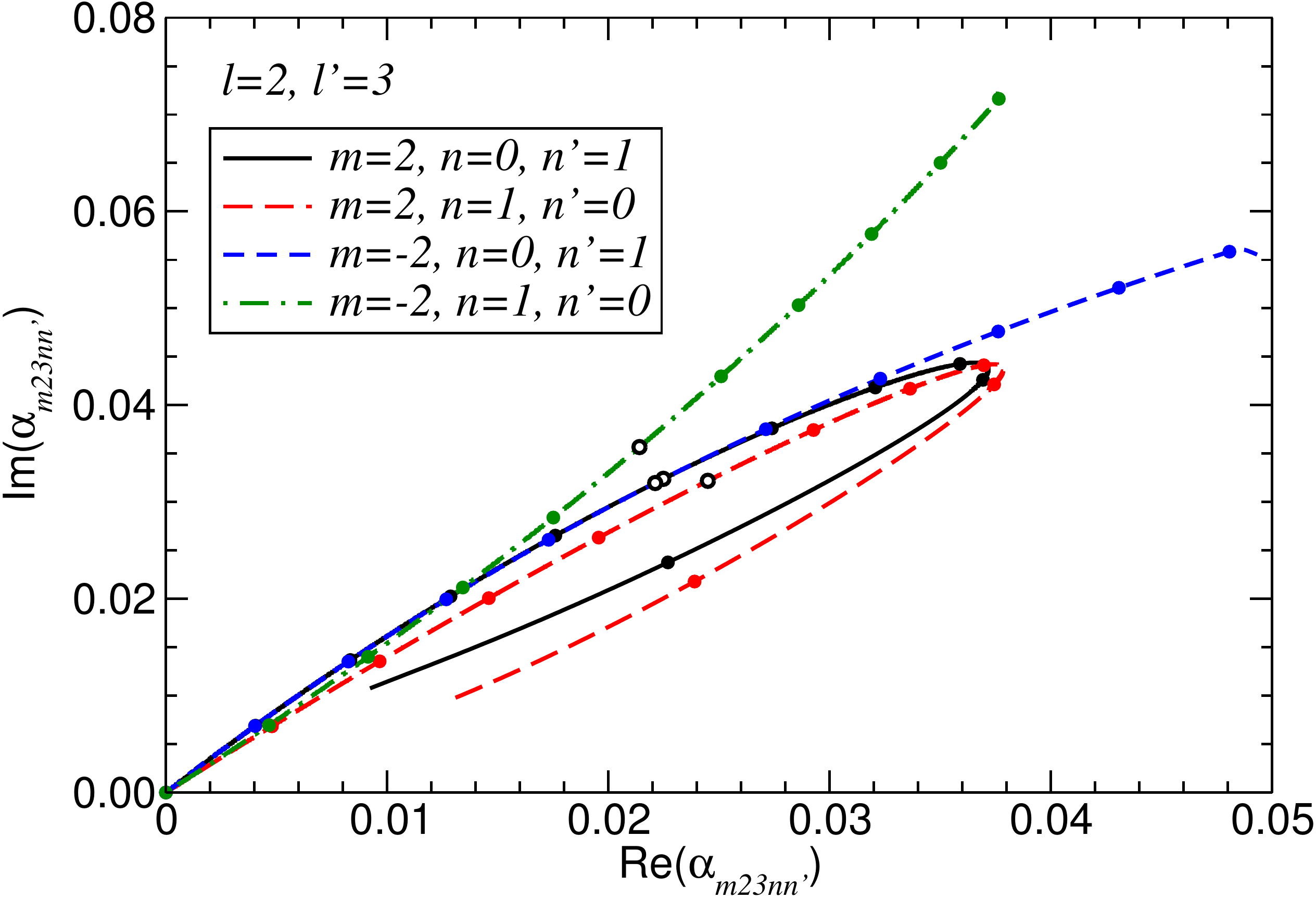,width=0.5\textwidth,angle=0,clip=true}\\
\end{tabular}
\caption{Trajectories described in the complex plane by the
  spheroidal-spheroidal mixing coefficients $\alpha_{m\ell
    \ell'nn'}(j)$ with: (1) $\ell=2$, $\ell'=3$, $n=n'=0$ and
  $-\ell\leq m\leq \ell$ (left panel); (2) $\ell=2$, $\ell'=3$,
  $|m|=2$, and either $(n,\,n')=(0,\,1)$ or $(n,\,n')=(1,\,0)$.
}
\label{fig:spheroidalsl2l3}
\end{center}
\end{figure*}

Motivated by the fact that ringdown waveforms should be expanded in
terms of SWSHs rather than spin-weighted spherical harmonics
\cite{Berti:2005ys}, Ref.~\cite{Berti:2005gp} carried out a limited
and preliminary investigation of the spheroidal-spheroidal mixing
coefficients.  Table I of \cite{Berti:2005gp} compared a numerical
calculation of selected spheroidal-spheroidal mixing coefficients
$\alpha_{m\ell \ell'nn'}(j)$, as defined in Eq.~(\ref{acoeff}), with
the Press-Teukolsky perturbation theory calculation. The constants
$\alpha_{m\ell \ell'nn'}(j)$ computed using Leaver's method were
listed in Tables~II and III of \cite{Berti:2005gp} for $j=0.8$ and
selected values of the indices.

Here we extend those preliminary calculations to generic values of $j$
and to all modes of relevance for gravitational-wave data analysis.
Representative results are shown in Figs.~\ref{fig:spheroidalsn07}
and~\ref{fig:spheroidalsl2l3}. Fig.~\ref{fig:spheroidalsn07} shows the
scalar product between the dominant mode in black-hole binary merger
simulations ($\ell=\ell'=m=2$, $n=0$) and higher overtones with the
same angular dependence (same $\ell=\ell'=m$). All modes describe
loops that begin and end close to $\alpha_{2220n'}=1$; the one
exception, as usual, is the QNM with $n'=5$.

The most relevant spheroidal-spheroidal mixing coefficients to
understand black-hole binary simulations are small-$n$ overtones with
low angular indices $(\ell,~\ell')$ equal to either $2$ or $3$. Some
of these mixing coefficients are plotted, with the usual conventions,
in Fig.~\ref{fig:spheroidalsl2l3}. In particular, we show (1) the
$m$-dependence of spheroidal-spheroidal overlaps when $\ell=2$,
$\ell'=3$, $n=n'=0$, and (2) the overlap between the fundamental mode
and the first overtone when $\ell=2$, $\ell'=3$ and $|m|=2$.

\subsection{Fitting formulas for the mixing coefficients\label{sec:fit}}

As illustrated in Fig.~\ref{fig:m2n0lowest}, we can reproduce the
numerical data for the mixing coefficients to satisfactory accuracy
(absolute deviations being typically smaller than $10^{-4}$ for the
dominant modes, and smaller than a few times $10^{-3}$ for all modes we
considered) with the following power-law fits:

\bea
{\rm Re}(\mu_{m\ell \ell'n'}) &=& \delta_{\ell \ell'} + p_1 j^{p_2} + p_3 j^{p_4}\,, \nn \\
{\rm Im}(\mu_{m\ell \ell'n'}) &=& q_1 j^{q_2} + q_3 j^{q_4}\,. \label{mufit}
\eea

Table~\ref{fitcoeffs} lists the fitting parameters $(p_i,\,q_i)$
($i=1,\dots\,,4$) for some combinations of $(m, \ell, \ell', n')$ that
are particularly relevant in black-hole binary mergers. These values
were chosen as particularly significant because 
\begin{itemize}
\item[(i)] Ref.~\cite{London:2014cma} successfully extracted QNMs with
  $(\ell,\,m,\,n)$=$(2,\,2,\,0)$, $(3,\,2,\,0)$ and $(2,\,2,\,1)$ from
  numerical simulations of comparable mass black-hole mergers, showing
  that mode mixing plays an important role in the extraction
  procedure; and

\item[(ii)] Ref.~\cite{Taracchini:2014zpa} pointed out that mode
  mixing plays an important role also for extreme mass-ratio binaries
  (see e.g. their Fig.~7). In addition, they found that negative-$m$,
  ``counterrotating'' modes (or ``mirror modes'': see
  \cite{Berti:2005ys} for a discussion) contribute to the mixing,
  because frame dragging can change the sign of the orbital frequency
  of the plunging particle. This finding was confirmed by more recent
  time-domain calculations \cite{Harms:2014dqa}.
\end{itemize}

Table~\ref{fitcoeffs} is only representative. Comprehensive tables
listing these fitting parameters for scalar, electromagnetic and
gravitational modes with $|s|\leq \ell \leq 7$, $-\ell \leq m \leq
\ell$, $-\ell \leq m' \leq \ell$, $0\leq n'\leq 7$, $s=-2$ are
publicly available online at \cite{rdweb}, where we also provide
fitting parameters for the $\alpha_{m\ell\ell'nn'}$'s.

\begin{table*}[!ht]
\begin{center}
\begin{tabular}{cccc|cccc|cccc} \hline
\multicolumn{4}{c|}{Indices}
&\multicolumn{4}{c|}{${\rm Re}(\mu_{m\ell \ell'n'})$}
&\multicolumn{4}{c}{${\rm Im}(\mu_{m\ell \ell'n'})$}
\\
 $m$ & $\ell$ & $\ell'$ & $n'$ & $10^5 p_1$ & $p_2$ & $10^5 p_3$ & $p_4$ & $10^5 q_1$ & $q_2$ & $10^5 q_3$ & $q_4$ \\\hline
 $2$ & $2$ & $2$ & $0$ & $-740$ & $2.889$ & $-661$ & $17.129$ & $1530$ & $1.219$ & $-934$ & $24.992$ \\
 $2$ & $2$ & $2$ & $1$ & $-873$ & $2.655$ & $-539$ & $15.665$ & $4573$ & $1.209$ & $-2801$ & $25.451$ \\
 $2$ & $2$ & $3$ & $0$ & $14095$ & $1.112$ & $4395$ & $6.144$ & $1323$ & $0.854$ & $-852$ & $7.042$ \\
 $2$ & $3$ & $2$ & $0$ & $-10351$ & $1.223$ & $-5750$ & $8.705$ & $-1600$ & $0.953$ & $1003$ & $14.755$ \\
 $-2$ & $2$ & $2$ & $0$ & $-1437$ & $2.118$ & $1035$ & $2.229$ & $-7015$ & $1.005$ & $67$ & $3.527$ \\
 $-2$ & $2$ & $2$ & $1$ & $-2659$ & $2.007$ & $53$ & $4.245$ & $-21809$ & $1.008$ & $221$ & $4.248$ \\
 $-2$ & $2$ & $3$ & $0$ & $14971$ & $1.048$ & $-5463$ & $1.358$ & $18467$ & $1.015$ & $-10753$ & $1.876$ \\
 $-2$ & $3$ & $2$ & $0$ & $-13475$ & $1.088$ & $7963$ & $1.279$ & $-1744$ & $1.011$ & $516$ & $1.821$ \\\hline
\end{tabular}
\caption{\label{fitcoeffs}Fitting function parameters in
  Eq.~\eqref{mufit} for some of the $\mu_{m\ell \ell' n'}$'s that are
  most relevant in black-hole binary modeling.
}
\end{center}
\end{table*}

\section{Conclusions}

This paper was mainly motivated by recent investigations of
spherical-spheroidal mode mixing in black-hole binary mergers
\cite{Buonanno:2006ui,Kelly:2012nd,London:2014cma,Taracchini:2014zpa}.
For this reason our analysis was limited to four-dimensional SWSHs and
low-order overtones. Despite these limitations, we expect the
``dictionary'' developed in this paper to be useful in several
applications of black hole perturbation theory, including the
construction of phenomenological models of black-hole mergers, studies
of Green's functions in black-hole backgrounds, self-force
investigations (see e.g.~\cite{Casals:2013mpa,Wardell:2014kea}) and
calculations of Hawking radiation.

It would be interesting to extend our work to higher overtones, that
may have some relation with black-hole area quantization (see
e.g. \cite{Berti:2003zu,Berti:2003jh,Neitzke:2003mz,Berti:2004um,Hod:2005ha,Keshet:2007nv,Keshet:2007be,Kao:2008sv}),
or \cite{Berti:2004md,Berti:2009kk} for reviews). It would also be
useful to investigate mixing coefficients for higher-dimensional
spheroidal harmonics, that are of interest for the phenomenology of
black-hole formation in high-energy particle collisions
\cite{Kanti:2004nr} and to assess the stability of higher-dimensional
rotating black holes
\cite{Frolov:2002xf,Ida:2002zk,Berti:2005gp,Kunduri:2006qa,Hoxha:2000jf,Dias:2012tq}.
Furthermore our analysis was limited to spin values that are not very
close to $j=1$, and it calls for a more careful investigation of the
nearly extremal regime, where a bifurcation of the spectrum can occur
\cite{Yang:2012pj,Yang:2013uba} and lead to turbulent behavior
\cite{Yang:2014tla}.

The numerical data and fitting coefficients computed in this paper are
publicly available for download \cite{rdweb}. The webpage includes
also spherical-spheroidal mixing coefficients for SWSHs with $s=-1$
and $s=0$, that were not reported in this paper because they are
qualitatively similar to the data for spin weight $s=-2$.

\begin{acknowledgments}
We are grateful to Michalis Agathos, Riccardo Sturani, Scott Hughes,
Alessandra Buonanno, Andrea Taracchini, Gaurav Khanna and Sebastiano
Bernuzzi for correspondence and conversations that stimulated our
interest in this problem. This research was supported by NSF CAREER
Grant No.~PHY-1055103.
\end{acknowledgments}

\appendix

\section{Perturbative evaluation of the mixing coefficients\label{sec:PT}}

As mentioned in the main text, the SWSH equation can be solved via an
expansion in powers of $c$ using standard perturbation
theory~\cite{Press:1973zz}. For $c=0$ the solutions are ordinary
spin-weighted spherical harmonics
\cite{Newman:1966ub,Goldberg:1966uu}. The next-order correction can be
found in Eq.~(3.7) of Ref.~\cite{Press:1973zz} (see also Appendix F of
\cite{Tagoshi:1996gh}); the result is Eq.~(\ref{pert-th}), where
\be
\langle s \ell' m|\mathfrak{h}_1|s \ell m\rangle=\int {}_sY^*_{\ell' m} \mathfrak{h}_1\, {}_sY_{\ell m} \, d\Omega,\,
\ee
\be
\mathfrak{h}_1\equiv (a\omega)^2 \cos^2\theta-2a\omega s \cos\theta\,.
\ee
The integral can be evaluated using the identities
\begin{align}
 &\langle s \ell' m | \cos\theta | s \ell m \rangle \nn \\
 &= \left(\frac{2\ell+1}{2\ell' + 1}\right)^{1/2} \langle \ell, 1, m, 0| \ell', m \rangle \langle \ell, 1, -s, 0| \ell', -s \rangle\,,\nn \\
 &\langle s \ell' m | \cos^2\theta | s \ell m \rangle = \frac{1}{3} \delta_{\ell,\ell'}\nonumber\\
 &+ \frac{2}{3}\left(\frac{2\ell+1}{2\ell' + 1}\right)^{1/2} \langle \ell, 2, m, 0| \ell', m \rangle \langle \ell, 2, -s, 0| \ell', -s \rangle\,,\nn
\end{align}
where $\langle \ell_1, \ell_2, m_1, m_2 | L, M\rangle$ is a
Clebsch-Gordan coefficient.


%

\end{document}